# Towards Usage-based Impact Metrics:

## First Results from the MESUR Project

Johan Bollen
Digital Library Research & Prototyping Team
Los Alamos National Laboratory
Los Alamos, NM 87545
jbollen@lanl.gov

Herbert Van de Sompel
Digital Library Research & Prototyping Team
Los Alamos National Laboratory
Los Alamos, NM 87545
herbertv@lanl.gov

Marko A. Rodriguez
T-7, Center for Non-Linear Studies
Los Alamos National Laboratory
Los Alamos, NM 87545
marko@lanl.gov

## ABSTRACT

Scholarly usage data holds the potential to be used as a tool to study the dynamics of scholarship in real time, and to form the basis for the definition of novel metrics of scholarly impact. However, the formal groundwork to reliably and validly exploit usage data is lacking, and the exact nature, meaning and applicability of usage-based metrics is poorly understood. The MESUR project funded by the Andrew W. Mellon Foundation constitutes a systematic effort to define, validate and cross-validate a range of usage-based metrics of scholarly impact. MESUR has collected nearly 1 billion usage events as well as all associated bibliographic and citation data from significant publishers, aggregators and institutional consortia to construct a large-scale usage data reference set. This paper describes some major challenges related to aggregating and processing usage data, and discusses preliminary results obtained from analyzing the MESUR reference data set. The results confirm the intrinsic value of scholarly usage data, and support the feasibility of reliable and valid usage-based metrics of scholarly impact.

## Categories and Subject Descriptors

I.2.4 [**Knowledge Representation Formalisms and Methods**]: Semantic Networks; H.2.8 [**Database Applications**]: Data mining; H.3.7 [**Digital Libraries**]: Collection

## 1. INTRODUCTION

Citation data are routinely used to assess the impact of journals, journal articles, scholarly authors, and the institutions these authors are affiliated with. Such assessments increasingly influence promotion decisions, science funding and public policy [6, 16, 18]. However, scholarly communication is a multi-phased process that is difficult to capture in citation statistics only. The scholarly cycle begins with the development of an idea, eventually resulting in its publication and subsequent citation. Citations only occur at the end of this cycle, but throughout the scholarly process articles are discovered, downloaded, e-mailed to peers, read and saved for later consultation. Usage events at all phases of the scholarly process are now commonly recorded by online scholarly information services. It is thus conceivable that usage data can provide new insights in the scholarly process, and that novel measures of scholarly impact can be derived from usage data.

In fact, usage data has some distinct advantages over citation data. Most importantly, usage data can be recorded as soon as a scholarly artifact is made available online. It therefore instantly reflects the dynamics of the scientific process, as opposed to citation data that is subject to significant publication delays [8] and therefore documents past developments. Also, usage data can be recorded for a wide variety of accessible scholarly artifacts, not just for journal articles, and it can capture the actions of a user group that significantly extends beyond the authors of journal articles responsible for citations.

Usage data thus holds the promise of greatly expanding the possibilities for scholarly assessment. Recent studies of specific usage data sets confirm this potential [13, 14]. However, scholarly usage data and usage-based impact metrics have not yet made inroads as reliable and community-accepted tools. The reasons for this lack of acceptance are manifold and include the absence of standards to record and exchange usage data, uncertainties related to the compilation of usage data sets that credibly represent global scholarly activity, and a limited understanding of the nature of usage based metrics as well of their exact meaning and applicability.

The MESUR[1] project conducted by the Digital Library Research & Prototyping Team at the Los Alamos National Laboratory (LANL) was funded by the Andrew W. Mellon Foundation to investigate the feasibility of impact metrics derived from usage data. The core activities of the project are as follows (Fig. 1):

1. **Creating a reference data set:** MESUR has invested significant energy to compile a large-scale col-

---

[1]Pronounced "measure", an acronym for "Metrics from Scholarly Usage of Resources".

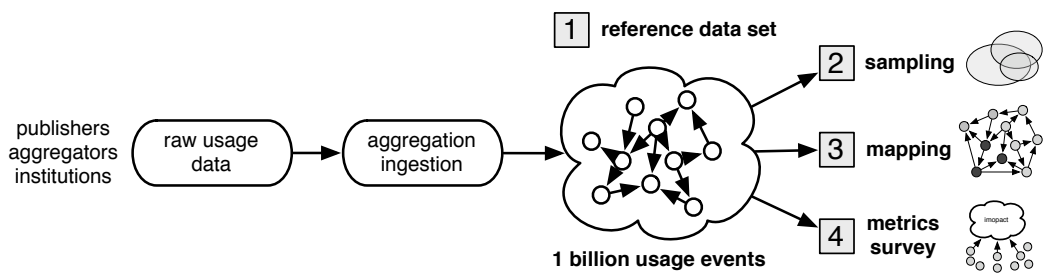

Figure 1: Overview of MESUR project phases.

lection of usage logs. Usage data, as well as associated bibliographic and citation information, was obtained from 14 significant publishers, aggregators and institutional consortia. The resulting *reference data set* serves as a "level playing field" for a subsequent survey of a wide range of usage-based metrics of impact.

2. **Researching sampling bias:** MESUR examines the effects of sampling biases on its reference data set to determine whether and how a usage data set can be compiled that is representative of global scholarly usage.

3. **Generating maps of science:** MESUR produces maps of science on the basis of its reference data set. These maps help assess the meaningfulness of usage data as such, indicate the validity of the collected usage data, and assist in investigating domain-specific effects on usage-based impact metrics.

4. **Defining and validating usage-based metrics:** MESUR defines a wide range of usage-based metrics, calculates them for the established reference data set, and assesses their validity and reliability. The final objective is to publish insights regarding the meaning and applicability of a set of usage-based metrics.

This article introduces preliminary results from the MESUR project, all of which strongly confirm the potential of scholarly usage data as a tool to study the dynamics of scholarship in real time, and to form the basis for the definition of novel metrics of scholarly impact. Section 2 describes the size, origin, and representation of the MESUR reference data set. Section 3 discusses initial findings in the realm of sample bias, and Section 4 shows the first ever map of science created on the basis of a substantial scholarly usage data set. Section 5 introduces a variety of impact metrics derived from both usage and citation data, and describes findings regarding their interrelation. Conclusions are presented in Section 6.

## 2. CREATING A REFERENCE DATA SET

A significant portion of MESUR's efforts in 2007 was invested in aggregating usage data from a variety of scholarly information services, and in establishing the necessary infrastructure to process, represent, store, and study the obtained usage data.

### 2.1 Present scale and span

So far, MESUR reached agreements for the exchange of usage data with 14 parties, and as a result has compiled a data set covering over **1 billion article-level usage events**, as well as all associated bibliographic and citation data. The breakdown of usage data sources is as follows[2]:

**Publishers** Six major international scholarly publishers.
**Aggregators** Four significant international aggregators.
**Institutions** Four large USA-based university consortia.

Further characteristics of the data set are as follows:

1. The usage events span nearly **5 years (2002-2007)** of activity, although not all data from the aforementioned contributors span the same time period.
2. The collected usage data spans more than **100,000 serials**, including scholarly journals, newspapers, etc.
3. The collected journal **citation data** spans about **10,000 journals and nearly 10 years**.
4. In addition to raw usage events, journal usage statistics have been collected in the form of **COUNTER reports** [21] that cover nearly **2000 institutions worldwide**.

With the exception of COUNTER reports, the obtained usage data was required to contain at least the following data fields: an anonymous session and/or user identifier, an article identifier, a date and time at which a request pertaining to the identified article took place, and an indication of the request type (e.g. article download, abstract view, etc.) As a result, it is possible to extract the various articles that users requested a service for in the course of a given session, and to reconstruct the clickstream of these users in the information system that recorded the usage data.

### 2.2 Filtering, parsing, and de-duplication

A significant amount of data processing must be performed to turn the heterogeneous usage data collections obtained from a variety of sources into a reference data set that provides a solid basis to perform cross-source analysis:

1. Anonymization: Understandably, privacy concerns are central to discussions with potential suppliers of usage data. Most agreements thus contain explicit statements with this regard. As a result, all usage data in the MESUR reference data set is anonymized both regarding individual and institutional identity. In certain cases, the usage data is provided by the source in an anonymized form, in other cases MESUR is responsible for the required processing.

---
[2]MESUR's agreements preclude the identification of its most significant contributors.

2. **Filtering:** In order to obtain a reference data set that only pertains to human usage of scholarly artifacts, MESUR detects and removes usage events generated by machine agents such as crawlers, spiders and bots. MESUR's statistical detection algorithms rely on a variety of variables such as session length, time delay between requests, and other parameters that indicate a non-human request origin.

3. **De-duplication:** Probably the greatest challenge in data processing is the clustering of usage events pertaining to the same scholarly artifact. Lacking unique identifiers that are truly used globally, and given the significant variances in bibliographic data to describe the same scholarly artifact across information systems, MESUR faces an extensive de-deduplication task. Its de-duplication approach leverages a high quality scholarly bibliographic collection describing over 50 million journal articles, and an extensive database listing equivalence classes for a variety of journal title and ISSN number variations. The de-duplication itself relies on a combination of ad-hoc heuristics and fuzzy matching approaches that statistically compare the metadata provided for a usage event to MESUR's bibliographic collection.

4. **Session-based grouping:** Usage data is typically recorded and hence provided to MESUR as a time-sequential list of individual events recorded by an information system; different events generated by the same agent in the course of a certain time span are not grouped. Since MESUR follows an approach of usage data analysis inspired by clickstream concepts [12, 11] grouping events is an essential processing sub-task that needs to be performed before ingesting the usage data into the reference data set. Session identifiers, anonymized user identifiers, anyonymized IP addresses, and event timestamps are information elements that are at the core of this process.

It should be noted that both the filtering and de-duplication sub-tasks are inherently statistical procedures, and that the achieved success rates influence the quality of the reference data set. Therefore, uncertainty quantification is important to MESUR as it will help to assess the reliability of results obtained from mining the reference data set. At the time of writing, a formal approach with this regard is being developed.

## 2.3 Reference data set representation

The requirement to handle a variety of semantic relationships (publishes, cites, uses) and different types of content (bibliographic data, citation data, usage data), led MESUR to define a context-centric OWL ontology that models the scholarly communication process [19][3]. This ontology forms the basis for the representation of the reference data set in the MESUR infrastructure. RDF [15] triple databases are the natural habitat for data represented in this manner, and they provide great flexibility for data analysis without the need for extensive upfront application design. However, their scalability and retrieval efficiency are generally not on a par with the most competitive relational database products. Therefore, the MESUR project uses a combination of a relational database to store and query item (e.g. article) metadata, and a triple database[4] to store and query semantic relationships among items. Unique identifiers for these items are shared among these storage infrastructures and allow jumping from one to the other as needed.

---
[3] http://www.mesur.org/schemas/2007-01/mesur

## 2.4 Research data set

The MESUR reference data now consists of 1 billion individual usage events that were recorded at the document-level and processed as described above. However, the time-consuming process of aggregation, filtering, parsing, and de-duplicating 1 billion usage events was terminated only recently. The preliminary results discussed in the following sections were generated on the basis of an early subset of the MESUR data set that was selected to offer the best possible outcomes at the time:

- **200 million article-level usage events:** A subset consisting of the most thoroughly validated and de-duplicated usage events.
- **Journal-level usage events:** All article-level usage events were converted to journal-level usage events to facilitate the interpretation and cross-validation of initial results.
- **All request types included:** Instead of making arbitrary determinations regarding the relative importance of various request types, all requests that are indicative of a user's interest in a given article are included. Multiple consecutive requests pertaining to the same article are conflated to one event. Future analysis will focus on determining which request types most validly represent user interest.

## 3. RESEARCHING SAMPLING BIAS

When usage data is recorded for a particular community, the conclusions that are drawn from its analysis apply only to that community. This was documented by Bollen (2008) [4] who shows that the rankings of a proposed Usage Impact Factor are strongly affected by the characteristics of the community for which usage data is recorded. As a result, one can assume that substantial usage data sets must be aggregated from a variety of sources in order to derive conclusions that have global reach [3]. This assumption seems to be confirmed by the pattern that emerges as the MESUR reference data set grows and becomes more diverse over time.

Table 1 lists the five highest-ranked journals according to their usage[5] at LANL, one of the initial usage data sets in the MESUR reference data set. The ranking is based on about 1.5 million usage events. As one can expect, the LANL ranking is dominated by physics journals, in particular those focused on material and nuclear science. The usage-based ranking deviates strongly from the citation-based Thomson Scientific journal Impact Factor (IF) [6] listed in the 3rd column.

A different picture emerges for the usage data obtained from the California State University (CSU) system. This data set was recorded by the OpenURL-compliant linking servers deployed on nine campuses of the CSU system, and comprises 3.5 million usage events. Table 2 provides the five highest-ranked journals according to their usage at CSU.

---
[4] http://agraph.franz.com/allegrograph/
[5] Usage PageRank as discussed in Section 5.
[6] Discussed in Section 5

| Usage | | IF (2003) | | |
|---|---|---|---|---|
| rank | value | rank | value | journal |
| 1 | 60.196 | 155 | 7.035 | PHYS REV LETT |
| 2 | 37.568 | 739 | 2.950 | J CHEM PHYS |
| 3 | 34.618 | 2616 | 1.179 | J NUCL MATER |
| 4 | 31.132 | 1183 | 2.202 | PHYS REV E |
| 5 | 30.441 | 1209 | 2.171 | J APPL PHYS |

Table 1: Usage-based rankings produced from 1M LANL usage events.

Clearly, the CSU community is significantly larger and more diverse than LANL. Interestingly enough, the usage-based ranking for CSU better approximates the IF, although the Journal of American Child Psychology, and the American Journal of Psychiatrics, ranked fourth and fifth respectively, clearly still reveal community bias, i.e. they have high usage within the CSU community but a comparatively low IF.

| Usage | | IF (2003) | | |
|---|---|---|---|---|
| rank | value | rank | value | journal |
| 1 | 78.565 | 22 | 21.455 | JAMA |
| 2 | 71.414 | 11 | 29.781 | SCIENCE |
| 3 | 60.373 | 8 | 30.979 | NATURE |
| 4 | 40.828 | 446 | 3.779 | JAM CHILD PSY |
| 5 | 39.708 | 152 | 7.157 | AM J PSYCHIAT |

Table 2: Usage-based rankings produced from 3.5M CSU usage events.

The contrast between the rankings derived from the aforementioned institution-specific data sets and those computed for the current MESUR research data set is striking. As mentioned, by the end of 2007, this data set consisted of 200 million usage events recorded by a variety of institutional linking servers, and online services operated by publishers and aggregators; this preliminary data set already spans a broad user community. Table 3 lists the resulting five highest-ranked journals; it indicates a strong convergence towards the IF, with the exception of the Lecture Notes on Computer Science (LNCS) which is nevertheless considered an important publication.

| Usage | | IF (2005) | | |
|---|---|---|---|---|
| rank | value | rank | value | journal |
| 1 | 15.830 | 6 | 30.927 | SCIENCE |
| 2 | 15.167 | 11 | 29.273 | NATURE |
| 3 | 12.798 | 89 | 10.231 | PNAS |
| 4 | 10.131 | 5989 | 0.402 | LNCS |
| 5 | 8.409 | 235 | 5.854 | J BIOL CHEM |

Table 3: Usage-based rankings produced from MESUR collection of 200M usage events.

The rankings listed in Tables 1, 2, and 3 illustrate two important considerations regarding usage data sampling. First, the characteristics of the community for which usage is recorded strongly shape usage-based impact rankings. Second, as the sample grows in size and scope, the preferences or biases of a particular community are leveled out, and an increasing convergence with the IF is observed. The observed convergence suggests that it is feasible to create a reference data set from which rankings with global reach can be derived. The authors are anxious to compute further rankings as the entire collection of 1 billion usage events is loaded, and to learn whether the convergence pattern persists.

## 4. GENERATING MAPS OF SCIENCE

The structural characteristics of MESUR's reference data set are carefully monitored as it increases in size and span. This section discusses the preliminary results of MESUR's efforts to produce detailed maps of science on the basis of journal networks derived from its usage and citation data. These maps provide an indication of the ability of MESUR's usage data to accurately model the structure of science.

### 4.1 Citation and usage networks

Journal networks can be created by connecting individual journals on the basis of a chosen relationship. Doing so is relatively straightforward for a citation relationship as citation databases such as Thomson Scientific's Journal Citation Records list the number of citations that point from one journal to another. Each row in the list represents a connection between a given pair of journals, and the number of citations indicates the strength of their connection. A journal citation network results when all connections are taken into account (Fig. 2). This approach has been extensively applied in other efforts to map science [7] on the basis of journal citation data.

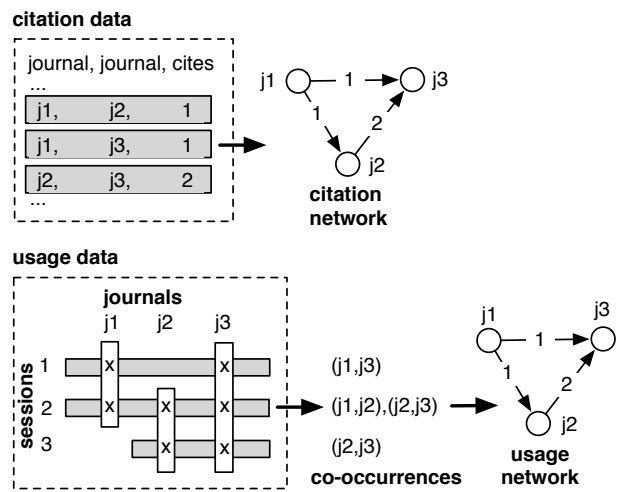

Figure 2: Building citation and usage networks.

Journal usage networks are created differently. Usage data is not expressed as a list of journal-to-journal connections, but as a flat, time-sequential list of article-level usage events from which journal connections must be derived. MESUR determines these journal connections using a process that is commonly employed by online recommender services such as Amazon.com and Netflix (i.e. "you may also be interested in these items"). The assumption at the basis of these systems is that the degree to which any pair of items is related is a function of the frequency by which users jointly purchase, download, or access them [20, 1]. This relationship is known as usage co-occurrence, and it is used to create MESUR's journal usage networks. However, since the MESUR usage data doesn't identify individual users, usage co-occurrence was reformulated in terms of sessions, indi-cated by anonymized session identifiers: the degree of relationship between any pair of journals is a function of the frequency by which they are jointly accessed within user sessions. Fig. 2 illustrates this process. Within a usage data set, usage events are grouped according to the session in which they occur. This allows determining how frequently a given pair of journals is accessed within the same session. This frequency determines the strength of the connection between this particular pair of journals. The connections thus extracted for each pair of journals can then be combined to form a journal usage network.

## 4.2 Mapping science from usage and citation data

The journal citation network was derived from Thomson Scientific's 2005 Journal Citation Records (Science and Social Science editions). Citations were limited to those that occurred in 2005 and pointed to articles published in 2004 and 2003 to ensure conformity of any computed metrics to the 2005 Impact Factor. The journal usage network was created on the basis of usage events recorded in 2006 extracted from the research data set described in Section 2.4 to limit the inevitable time difference between presently available citation and usage data.

Table 4 compares the data sets from which both networks were derived and shows the resulting network parameters. We note that the usage network was based on nearly 200 million usage events recorded for a total of more than 100,000 journals, whereas the citation network was based on a total of nearly 1.1 million citations recorded for 8,408 journals, a subset of all published journals selected by Thomson Scientific. Hence, the usage network is based on a much larger sample (200x) that pertains to a much larger set of journals (16x).

| Parameter | Usage | Citation |
|---|---:|---:|
| Source | MESUR | JCR |
| Date range | 2006 | 2005 |
| Data points | 200,000,000 | 1,082,313 |
| Journals | +100,000 | 8,408 |
| Connections | 6,933,787 | 1,082,313 |

Table 4: Overview of data from which MESUR's usage and citation networks were created.

The structural features of MESUR's usage and citation networks were compared in terms of:

**Network visualization:** a visualization aimed at presenting the most prominent journals and clusters of journals.

**Graph-theoretical parameters:** network density, centralization, and hierarchy.

Both usage and citation networks can not be visualized in their entirety due their large number of journals and connections. Therefore, Fig. 3 and Fig. 4 display a relevant subset of all journals and connections. This subset is selected as follows. First, all connections are ranked according to their connection strength (i.e. the number of citations or usage co-occurrences), and then only the top 5,000 connections are selected. Next, for each remaining journal, a maximum of 12 connections is shown. In addition, the visualization only includes journals that are part of the network's Largest Connected Component, which is the largest possible sub-network in which every journal is directly or indirectly connected to every other journal. This prevents the maps to be cluttered with small "island" networks. The remaining network is then graphically layed-out according to the Fruchterman-Reingold heuristic which uses "force-directed" placement to position connected journals in each other's proximity and minimize connection crossings [9]. The maps show only the titles of the most central journals within a given cluster to further reduce clutter. The radius of the circles in the maps is given by the natural logarithm of the number of connections for the journal. Journals with few connections thus have smaller circles.

| Network parameter | usage | citation |
|---|---:|---:|
| Connections | 5,000 | 5,000 |
| Max connection strength | 22,548 | 1,944 |
| Min connection strength (>0) | 62.5 | 97 |
| Centralization | 21.009 | 14.288 |
| Hierarchy | 0.0012 | 0.0018 |
| Density | 1.329 | 0.607 |

Table 5: Network parameters for usage and citation networks.

The visualizations shown are stable across repeated runs of the lay-out algorithm. A visual assessment of the generated maps as well as the network parameters in Table 5 reveal stark differences between the structure of the usage and citation networks, even though both were generated according to the same methodology. The usage-based map is one of many interlocking, interdisciplinary cycles and epicycles. The main ring connects (clock-wise) physics, chemistry, bio-sciences, environmental sciences, nutrition and agriculture, health care, and material sciences. Throughout this main ring off-shoots can be observed towards peripheral rings that connect various domains and applied subjects such as dermatology, livestock, alternative energy science and resource management. Of particular noteworthiness is the cluster containing psychology, cognitive science and education journals in the top-left section of the journal usage network. Also, there is a ring connecting psychology, urban studies, remote sensing, geology, material sciences, physics, medicine, neuro-imaging, neuroscience, and finally reaches back to psychology. Similarly, a ring connects chemistry to pharmacology, health-care (with an off-shoot to allergy and asthma research), to agriculture, bio-medicine and back to pharmacology. The latter sits appropriately between chemistry and health-care. Health-care itself is positioned between psychiatrics, rehabilitation, and nutrition.

The network-parameters in Table 5 [23] indicate that the journal usage network is denser ("Density"), less hierarchical ("Hierarchy") and more centralized ("Centralization"), i.e. less disjoint, than the journal citation network. The latter is characterized by a stronger decentralization, stronger hierarchical separation of the individual scholarly domains, and a lesser focus on the social sciences that are also less connected to the other sections of the graph. The journal citation network does however more strongly include Computer Science (bottom).

These results indicate that equally valid, rich and definitely more current maps of science can be generated from usage-data than from citation data.

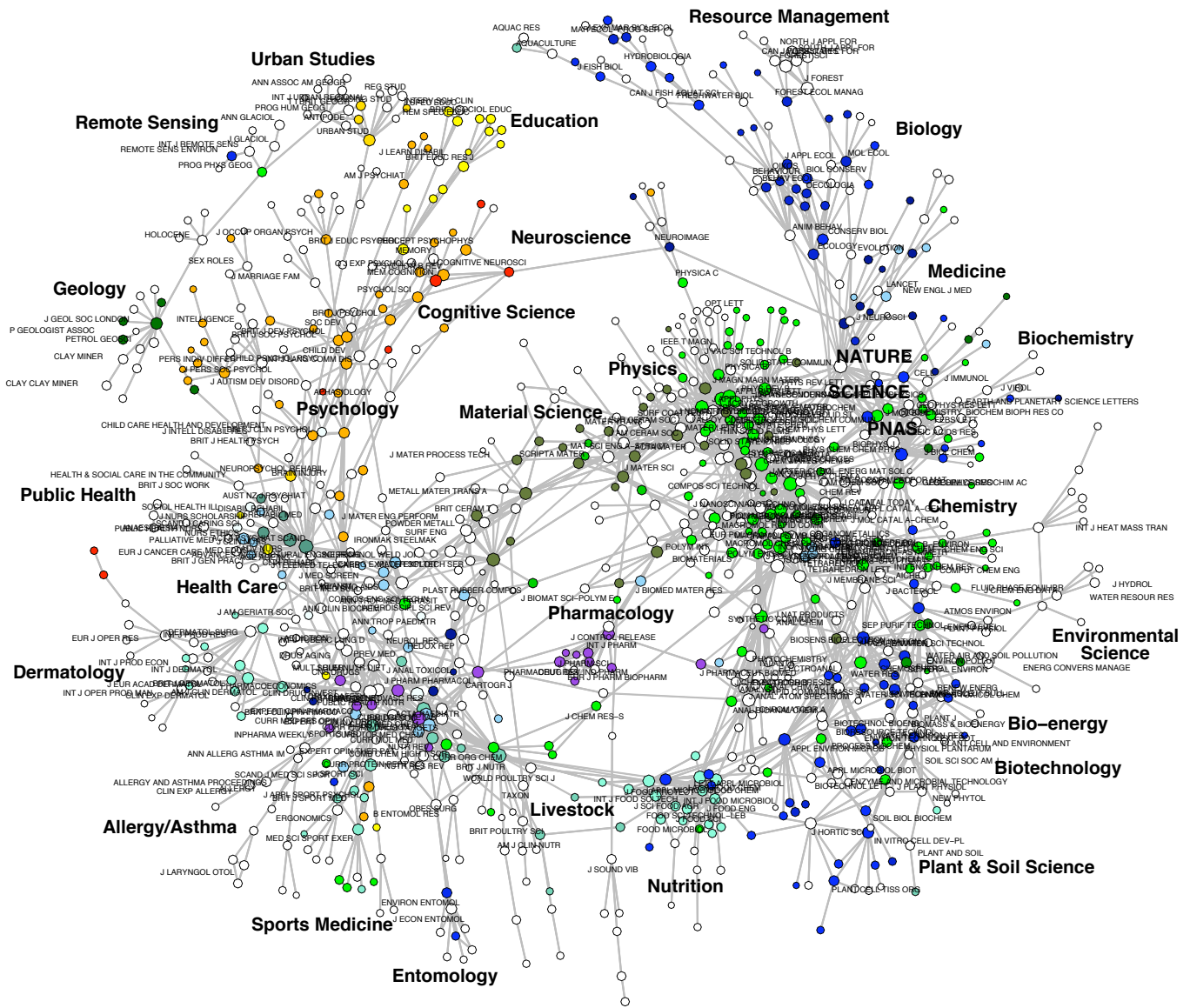

Figure 3: Visualization of usage network created from MESUR's 200M usage events.

## 5. USAGE-BASED METRICS

The journal usage and citation networks also enable the calculation of a variety of impact metrics. A total of 47 possible impact metrics were calculated, and the resulting rankings were analyzed to determine the degree to which usage- and citation-based metrics express similar or dissimilar aspect of scholarly impact.

### 5.1 Defining and validating usage-based metrics

The most common indicator of journal status is Thomson Scientific's journal Impact Factor (IF) that is published every year for a set of about 8,000 selected journals. The IF is defined as the average citation rate for articles published in a particular journal. A similar statistical approach to journal ranking has been proposed for journal usage data on the basis of COUNTER reports, i.e. the average amount of usage recorded for the articles published in a journal [4, 22].

However, to use a social analogy, one's importance is not solely assessed on the basis of *how many* people one knows. *Who* one knows and *how* one is embedded in a network of social relationships are equally important factors. Network theory has produced a rich literature on indicators to determine different facets of a person's status (e.g. prestige, popularity, trust) on the basis of social network structure, instead of using simple counts of the number of the person's relationships. Many of these indicators have found applications in other domains. For example, the Google search engine uses the PageRank metric to rank web pages on the basis of the WWW's hyperlink network structure. In addition, recent proposals have been made to rank journals according to their citation PageRank [2] and a range of social network indicators [23] calculated on the basis of citation data [5]. The generated usage network thus enables the calculation of a variety of structural, network-based impact metrics, such as PageRank, from usage data [4, 22]. MESUR has calculated an initial set of such metrics that can be classified according to the facets of scholarly impact they express (Table. 6). The abbreviation for each metric, used in following sections, is provided as well[7].

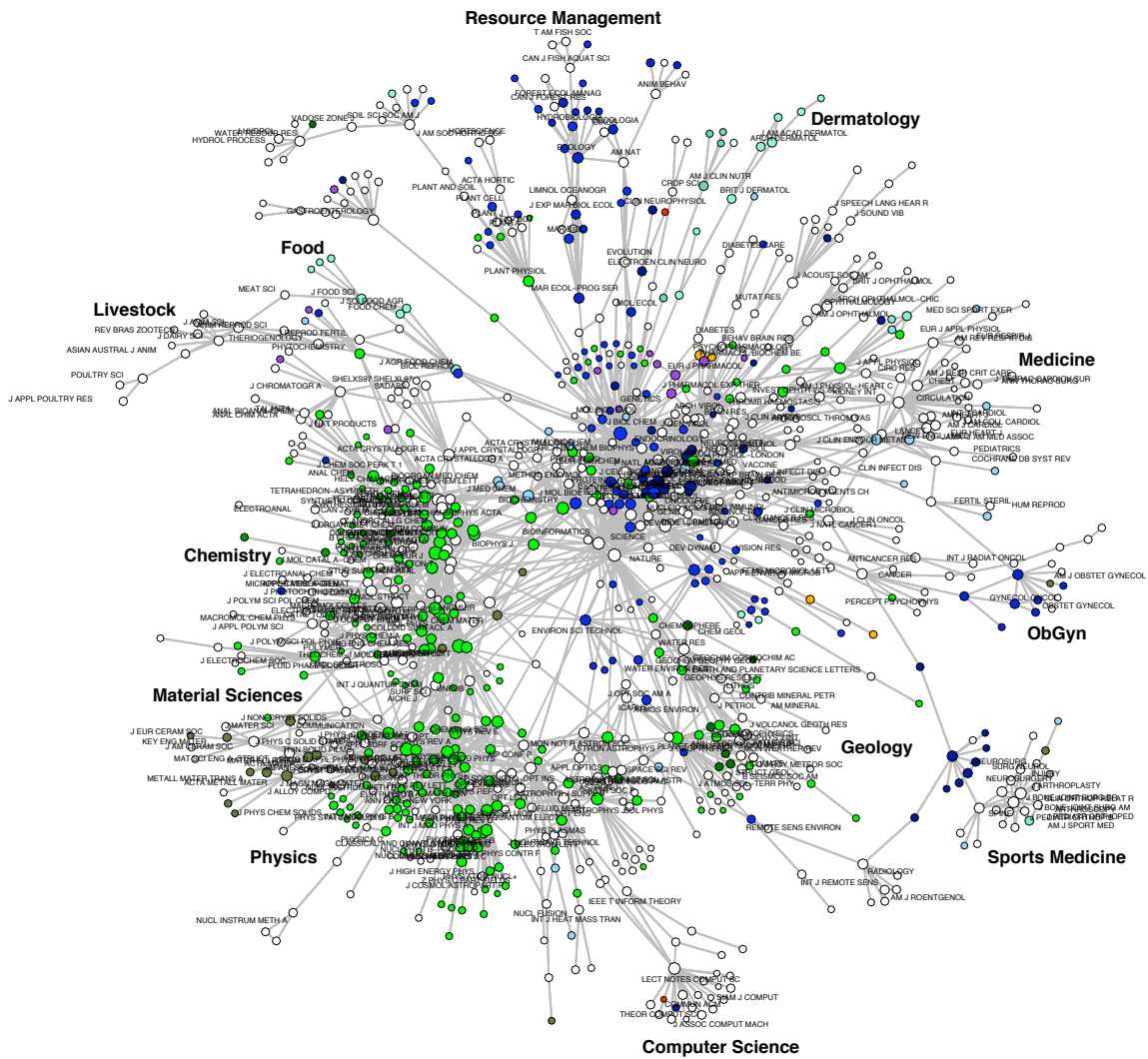

Figure 4: Visualization of citation network created from 2005 Journal Citation Records.

Each of the 9 listed metrics can be calculated for both the generated journal usage and journal citation networks. In addition, some can be defined to either take into account connection strengths or not. Some metrics can also be either normalized by the size of the network's largest connected components or not. These permutations lead to a total of *23 usage-based metrics* and *23 citation-based metrics*. With the addition of the Thomson Scientific journal Impact Factor a set of 47 metrics of scholarly impact result.

---

[7]The MESUR website offers detailed information on metric definitions and abbreviations: http://www.mesur.org/

## 5.2 Resulting rankings

All 47 metrics are calculated to investigate the possibility of reliable and valid usage-based metrics by comparing their rankings to citation-based rankings for the same set of journals, in particular the Thomson Scientific journal Impact Factor.

A sample of the produced rankings is shown in Table 7. These preliminary rankings demonstrate the ability of network-based metrics to express different facets of scholarly impact. Whereas the 2005 Impact Factor rankings are dominated by review journals in the medical domain (most often cited), the network-based metrics clearly manage to distinguish the most prestigious journals. Interestingly, the Lecture Notes in Computer Science and Lancet perform well in the citation-based betweenness rankings, indicating they frequently act as a bridge between other journals. Both journals also rank well in all three usage-based rankings. The latter overall approximate the network-based citation metrics quite closely. In fact, the rankings produced by all network-based metrics, whether citation-based or usage-

| Facet of impact | Metrics | Definition | Abbreviation |
|---|---|---|---|
| **Popularity** | In-Degree and | Number of links to- and from journal | ID |
| | Out-degree centrality, | | OD |
| | Journal Impact Factor | | IF |
| **Dispersion** | In-Degree entropy | Distribution of links to- and from journals | IE |
| | Out-degree entropy | | OE |
| **Friend-of-a-friend** | Closeness centrality | Average length of shortest path between journals | CL |
| **Power** | Betweenness centrality | Number of times journals sits on shortest paths | BW |
| | Newman's load [17] | | NM |
| **Prestige** | PageRank | Number of prestigious journals that link to journal | PR, PG |

Table 6: Facets of status expressed by set of network-based metrics used by MESUR.

**Citation-based metrics**

| | 2005 Impact Factor | | PageRank x 100 | | Betweenness | |
|---|---|---|---|---|---|---|
| rank | value | journal | value | journal | value | journal |
| 1 | 49.794 | CA-CANCER J CLIN | 1.200 | SCIENCE | 0.076 | PNAS |
| 2 | 47.400 | ANNU REV IMMUNOL | 1.100 | J BIOL CHEM | 0.072 | SCIENCE |
| 3 | 44.016 | NEW ENGL J MED | 1.100 | NATURE | 0.059 | NATURE |
| 4 | 33.456 | ANNU REV BIOCHEM | 1.100 | PNAS | 0.039 | LNCS |
| 5 | 31.694 | NAT REV CANCER | 0.600 | PHYS REV L | 0.017 | LANCET |

**Usage-based metrics**

| | In-degree | | PageRank x 100 | | Betweenness | |
|---|---|---|---|---|---|---|
| rank | value | journal | value | journal | value | journal |
| 1 | 4195 | SCIENCE | 0.160 | SCIENCE | 0.035 | SCIENCE |
| 2 | 4019 | NATURE | 0.150 | NATURE | 0.032 | NATURE |
| 3 | 3562 | PNAS | 0.130 | PNAS | 0.020 | PNAS |
| 4 | 2438 | J BIOL CHEM | 0.100 | LNCS | 0.017 | LNCS |
| 5 | 2432 | LNCS | 0.080 | J BIOL CHEM | 0.006 | LANCET |

Table 7: Rankings resulting from selected metrics

based, converge with the notable exception of the IF, which is not a network-based but rather a statistical metric.

## 5.3 The structure of metric correlations

The rankings listed in Table 7 are limited to the 5 highest ranked journals out of nearly 8,500. As such, they do not reflect the full degree of similarities and dissimilarities that exist between the investigated metrics. To quantitatively assess the differences between the 47 metrics, Spearman rank-order correlation coefficients are calculated between each pair of metric rankings, leading to a $47 \times 47$ matrix of metric correlations. This correlation matrix is denoted $C$, and each entry $c_{i,j} \in C$ indicates the Spearman rank-order correlation coefficient between metric $i$ and metric $j$ and ranges between $[-1, +1]$. For example, the following $4 \times 4$ sub-matrix shows the correlation values between the IF, citation betweenness, citation closeness, and usage betweenness:

$$\begin{pmatrix} 1.00 & 0.51 & 0.57 & 0.40 \\ 0.51 & 1.00 & 0.47 & 0.71 \\ 0.57 & 0.47 & 1.00 & 0.44 \\ 0.40 & 0.71 & 0.44 & 1.00 \end{pmatrix}$$

The correlation matrix $C$ can be used to map the similarities and dissimilarities between the various metrics using a Principal Component Analysis (PCA) [10]. A PCA determines the set of "dominant" eigenvectors, i.e. those with the highest eigenvalues, for the correlation (or co-variance) matrix between a set of variables. These original correlations are then mapped into the space spanned by the $k$ eigenvectors with the highest eigenvalues, the latter referred to as the principal components. A PCA that uses only the first 2 principal components of matrix $C$ will thus result in a 2D map that positions the metrics such that their distances in the map best reflect their original correlations. Such a map can be used to visually assess the similarities and dissimilarities between groups of metrics and to identify groups of metrics that indicate distinct facets of scholarly impact.

The 2D map resulting from a PCA of the metric correlation matrix $C$ is shown in Fig. 5. The first principal component (PCA1) forms the x-axis of the graph and represents about 53% of the variation in metric correlations. The second principal component (PCA2) forms the y-axis and represents about 27% of the variation in metric correlations. Combined, PCA1 and PCA2 thus capture about 80% of the variation in the metric correlations in matrix $C$ and therefore provide a good model of the original metric similarities shown in Fig. 5.

Each metric is labeled according to the abbreviations listed in Table 6, including an indication of whether it is citation- or usage-base ("CITE" or "USES"), takes into account connection weights ("W"), and has not been normalized to the size of the largest connected component ("UN"). The darkness of the background color indicates the density with which metrics cluster in a particular area, i.e. darker means denser. The map reveals a number of surprising features of the relationships between usage- and citation-based metrics. The x-axis, given by the first principal component (PCA1), represents the largest amount of variation in metric correlations. As expected it separates the metrics along the most primary distinction of their being either citation-based (right) or usage-based (left). The y-axis, given by the second principal component (PCA2), covers a lesser amount of variation in metric correlations and thus represents a more subtle, secondary distinction. Along the y-axis we find a vertical separation of citation metrics that ranges from closeness centrality on the top to degree centrality, including the IF in the middle, to a cluster of PageRank and betweenness centrality in the bottom range. A similar but less salient arrangement applies to the cluster of usage-based metrics. The vertical dimension (PCA2) can thus be interpreted as a gradient from "popularity" to "prestige". The former is indicated by closeness centrality, degree-centrality, and the IF. The latter is indicated by PageRank and betweenness centrality.

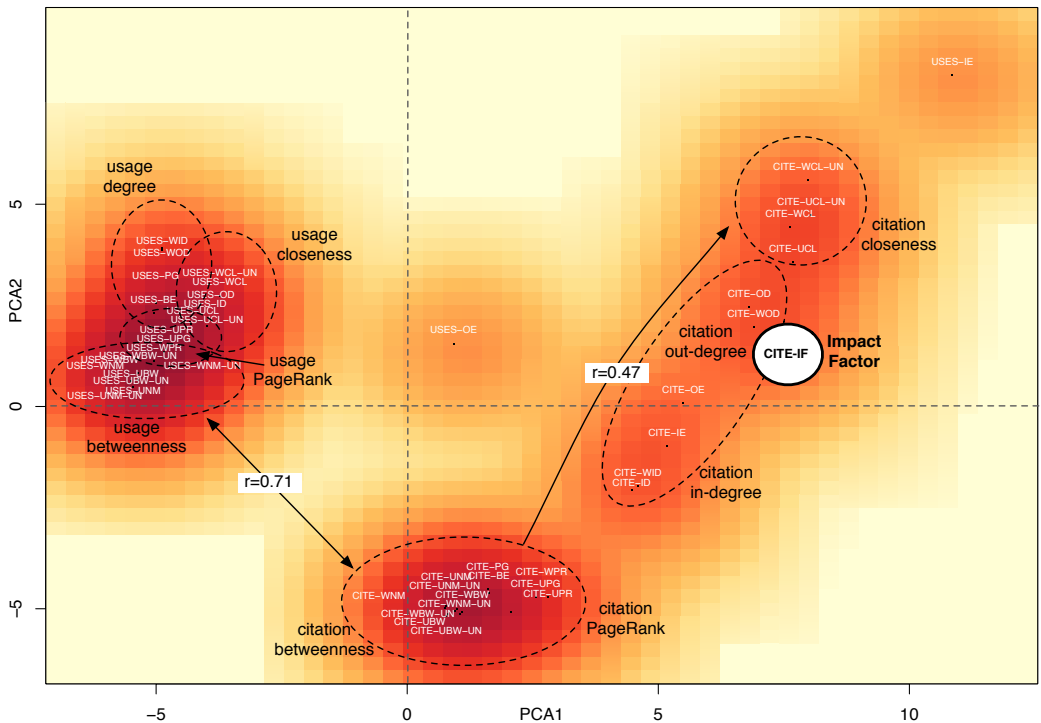

Figure 5: Principal component analysis of Spearman rank-order correlations between 47 preliminary MESUR metrics.

The map indicates a much stronger degree of agreement between the usage-based metrics than between the citation-based metrics, which may be caused by the significantly greater density of the journal usage network. The greater degree of "corroborating" data points in the usage graph may in fact explain the stronger convergence of the usage-based metrics.

Although the PCA map seems to strongly separate usage-from citation-based metrics, some usage-based metrics approximate certain citation-based metrics better than some citation-based metrics approximate one another. In fact, the Spearman rank-order correlation coefficient between usage-based betweenness centrality (middle, left) and citation-based betweenness centrality (bottom, middle) is 0.71 ($p < 0.001$) indicating they represent a similar aspect of scholarly status. However, the correlation between the same citation-based betweenness centrality (bottom, middle) and the citation-based closeness centrality (top, right) is only 0.47 ($p < 0.001$). Also, the citation-based IF differs more strongly from citation-based betweenness than citation-based betweenness differs from usage-based betweenness. In fact, the IF sits relatively isolated in a cluster of metrics that seems to express a different aspect of status than the cluster of all usage-based metrics, or the cluster that combines citation betweenness and citation PageRank.

These PCA results constitute only a preliminary, proof-of-concept analysis executed on the basis of a limited set of possible metrics. Nevertheless, they provide useful insights regarding the nature and interrelation of a set of common, plausible metrics of impact, both usage- and citation-based. As the MESUR reference data set expands and the set of investigated metrics grows, a more complete survey of usage- and citation-based metrics should result.

## 6. CONCLUSION

Usage-based metrics for scholarly assessment are not commonly accepted in spite of their considerable potential. This can be attributed to a number of factors described in the Introduction, but central to them all is the lack of real understanding of both usage data and usage-based metrics. Indeed, most existing research into usage-based metrics of scholarly impact focuses on single metrics whose characteristics are explored on the basis of usage data that has been recorded for particular scholarly communities. The MESUR project attempts to fundamentally increase our understanding of usage data. It aims to pave the way for an inclusion of usage-based metrics into the toolset used for the assessment of scholarly impact and move the domain beyond the long-established and often disputed IF. To that end, MESUR has established the largest existing reference data set of usage, citation and bibliographic data obtained from a variety of the world's most significant publishers, aggregators and institutional consortia. This reference data set forms the basis for a program aimed at the identification, validation and characterization of a range of usage-based metrics. This paper has described preliminary results derived from an analysis of a subset of the MESUR reference data set that consists of over 200 million article-level usage events. The results strongly point towards the imminent feasibility of usage-based metrics of impact. The created journal usage network surpasses the journal citation network in terms of the number of data points from which it has been generated, the density of its connections, the richness of its structure, and its ability to model the many interdependencies that characterize modern, interdisciplinary science. The rankings according to a set of proposed usage-based metrics correlate significantly with several citation-based metrics of impact, including Thomson Scientific's journal Impact Factor, but most strongly with network-based citation metrics that express prestige, including betweenness centrality and PageRank. A Principal Component Analysis of metric correlations confirms this pattern and further elucidates the major dimensions along which usage- and citation-based metrics express different facets of impact. This result strongly suggests that usage-based impact rankings may further converge as MESUR ingests its entire collection of 1 billion usage events, but that this convergence may very well be towards a notion of scholarly prestige different than the one expressed by the IF.

These preliminary results already provide a tantalizing preview of the possibility of usage-based metrics of impact that are more adaptive, more timely and more accurate than any other assessment metric that is presently available.

## 7. ACKNOWLEDGMENTS

This research is supported by a grant from the Andrew W. Mellon Foundation.